# A GEN AI Framework for Medical Note Generation

1st *Hui Yi, Leong
*Data Science Institute,*
*University of Chicago, Chicago, Illinois, USA.*

2nd Yi Fan, Gao
*Data Science Institute,*
*University of Chicago, Chicago, Illinois, USA.*

3rd Shuai, Ji
*Data Science Institute,*
*University of Chicago, Chicago, Illinois, USA.*

4th Bora Kalaycioglu
*University of Chicago Medical,*
*University of Chicago, Chicago, Illinois, USA.*

5th Uktu Pamuksuz
*Data Science Institute,*
*University of Chicago, Chicago, Illinois, USA.*

***Abstract*— The increasing administrative burden of medical documentation, particularly through Electronic Health Records (EHR), significantly reduces the time available for direct patient care and contributes to physician burnout. To address this issue, we propose MediNotes, an advanced generative AI framework designed to automate the creation of SOAP (Subjective, Objective, Assessment, Plan) notes from medical conversations. MediNotes integrates Large Language Models (LLMs), Retrieval-Augmented Generation (RAG), and Automatic Speech Recognition (ASR) to capture and process both text and voice inputs in real time or from recorded audio, generating structured and contextually accurate medical notes. The framework also incorporates advanced techniques like Quantized Low-Rank Adaptation (QLoRA) and Parameter-Efficient Fine-Tuning (PEFT) for efficient model fine-tuning in resource-constrained environments. Additionally, MediNotes offers a query-based retrieval system, allowing healthcare providers and patients to access relevant medical information quickly and accurately. Evaluations using the ACI-BENCH dataset demonstrate that MediNotes significantly improves the accuracy, efficiency, and usability of automated medical documentation, offering a robust solution to reduce the administrative burden on healthcare professionals while improving the quality of clinical workflows.**

## I. INTRODUCTION

Despite the widespread adoption of Electronic Health Records (EHRs), clinicians are increasingly overwhelmed by the management of vast amounts of medical data. This overload can lead to errors and negatively impact the quality of healthcare delivery, especially when the data is incorrect, incomplete, or irrelevant. Physicians are particularly burdened by administrative tasks, with documentation consuming between 25% and 50% of their time [1]. The extensive effort required for collecting, processing, and documenting dialogue data significantly reduces the time available for patient care, education, and clinical research [2].

Artificial Intelligence (AI), particularly Large Language Models (LLMs), is progressively transforming medical practice by offering innovative solutions to these challenges. LLMs have been utilized in healthcare for applications such as clinical decision support, medical writing, and patient interaction assistance [3]. However, their application in medical dialogue summarization remains underexplored. Given their capacity to process large volumes of unstructured data and generate coherent text, LLMs hold significant potential for tasks like summarizing patient information and assisting with medical documentation [3].

In this research, we continue previous study of fine-tuned model automated medical documentation [4], we introduce MediNotes, a generative AI framework designed to alleviate the documentation burden on healthcare providers. from what we the first framework Leveraging advanced speech recognition technologies and LLMs, MediNotes aims to generate SOAP (Subjective, Objective, Assessment, and Plan) notes from medical conversations in real time. The framework employs ambient listening to record interactions and automatically transcribes them into structured medical notes, thereby significantly reducing the time physicians spend on documentation. To ensure high-quality outputs, state-of-the-art Natural Language Processing (NLP) techniques are integrated to enhance accuracy and relevance.

Furthermore, MediNotes incorporates a user-friendly chatbot that enables patients and healthcare professionals to access relevant medical information quickly and efficiently. Utilizing a vector database and Retrieval-Augmented Generation (RAG), the system performs contextual searches to provide precise and appropriate information. By streamlining these processes, the framework aims to improve patient care and enhance the overall efficiency of healthcare delivery.

## II. RELATED WORK

The administrative burden on physicians is a critical issue in healthcare, with documentation tasks consuming up to 50% of their time [1]. This significant allocation detracts from patient care and contributes to physician burnout. Consequently, there is a pressing need for automation in clinical documentation to alleviate this workload.

Early attempts to automate this process involved statistical machine translation systems designed to convert patient-doctor interactions into written records. While innovative, these systems often failed to capture the complexity and nuances of medical discourse, leading to unreliable documentation [5]. The introduction of transformer-based models marked a significant advancement in natural language processing (NLP); however, their high computational requirements posed challenges for clinical settings.

** Corresponding author: yuki.leong@uchicago.edu*

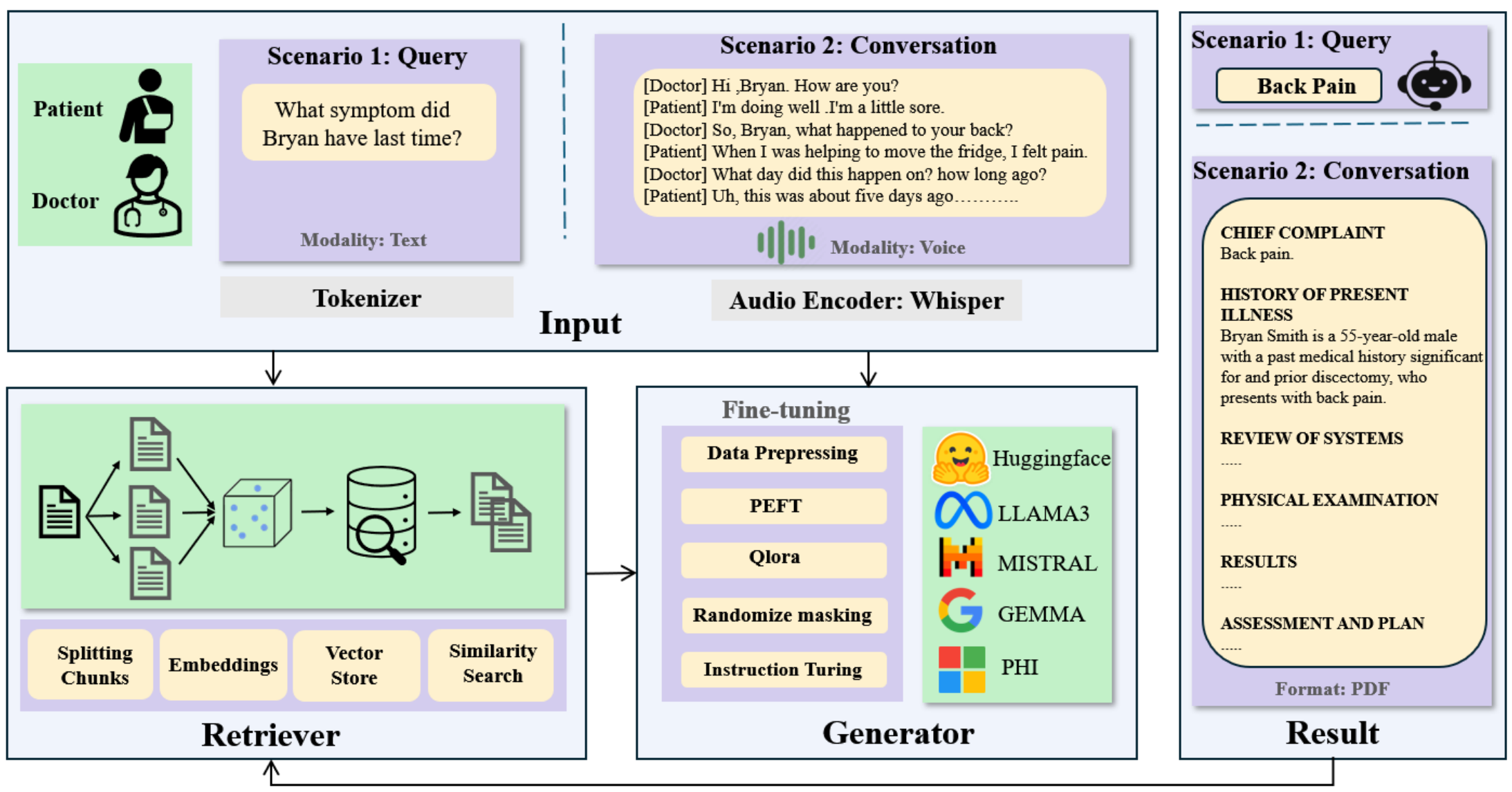

Fig. 1. Overview of the MediNotes Generative AI Framework for Medical Documentation. This figure demonstrates the workflow of the MediNotes framework, showcasing how it processes both text and voice inputs to generate structured medical documentation and data retrieval using advanced AI models and tech.

To address these challenges, recent research has explored methods to reduce computational demands without compromising performance. Leong et al. [4] proposed the use of Low-Rank Adaptation (LoRA), Parameter-Efficient Fine-Tuning (PEFT), and instruction fine-tuning techniques to efficiently fine-tune Large Language Models (LLMs) for automated medical documentation in resource-constrained environments. While this approach enhances computational efficiency, limitations remain. The accuracy of the generated documentation can still be improved, and these LLMs convert notes but lack the ability to memorize or update with the latest information. Additionally, they process text input exclusively and cannot interpret voice data, restricting their utility in real-time clinical settings. Therefore, incorporating additional functionalities is necessary to enhance the documentation process.

Beyond physician workload, patient recall of medical information is a significant concern. Studies have shown that patients forget 40–80% of the medical information provided by healthcare practitioners immediately after consultations [6]. This lapse negatively impacts patient outcomes and adherence to treatment plans. To assist patients in retrieving medical information, LLMs need to access external sources. Retrieval-Augmented Generation (RAG) has emerged as a promising technique in this context. RAG often combine pre-trained sequence-to-sequence models which is LLM with dense vector indexes of external databases, accessed through neural retrievers [7]. Thus, it enhances the accuracy and factuality by retrieving relevant data during text generation, thereby aiding both patients and healthcare providers

Even with advancements in automation and retrieval, challenges persist. The process of generating and retrieving information still requires manual input; users need to type queries into LLMs to generate responses. This requirement is particularly burdensome given the sheer volume of clinical notes. Data indicates that physicians generated 104 million notes for 1.9 million unique patients, totaling approximately 33 billion words in 6 years [8]. To alleviate some of this burden, the medical industry has increasingly turned to ambient listening and Automatic Speech Recognition (ASR) technology. ASR facilitates the automatic transcription of spoken language into text, enabling real-time documentation of patient-doctor interactions without manual input [9]. However, integrating ASR with LLMs to create a seamless, voice-activated documentation system remains an area requiring further exploration.

In addition to the prior works, other relevant research on automation for medical note generation and retrieval has made notable contributions, such as [10] and [11]. MedKnowts integrates a note-taking editor into Electronic Health Records (EHR) systems with information retrieval functionalities, streamlining clinical documentation and reducing the cognitive load associated with accessing patient data. However, despite its advantages, MedKnowts still requires significant manual input, relying on users to interact with the note-taking editor, which limits its efficiency in high-volume clinical settings that demand real-time documentation [10]. Similarly, [11] proposes natural language processing (NLP) models for aligning clinical dialogue with notes and summarizing patient visits. While this approach demonstrates potential for improving documentation, it primarily focuses on sentence alignment, restricting its adaptability to complex and dynamic medical conversations. Furthermore, it lacks real-time interaction capabilities, limiting its practical application in fast-paced clinical environments.

In contrast, our proposed solution—MediNotes—from what we understand the first framework study offers a more advanced AI-driven approach by integrating Large Language Models (LLMs), Retrieval-Augmented Generation (RAG), and Automatic Speech Recognition (ASR) technologies for medical note generation. By leveraging ambient listening, real-time voice transcription, and context-aware retrieval of medical information, MediNotes addresses the limitations of manual input and static summarization seen in previous works. This comprehensive AI solution not only reduces the administrative burden on physicians but also enhances the accuracy and efficiency of medical note generation by operating in real-time, thereby providing more advanced support in high-stakes healthcare environments.

## III. Methodology

The proposed MediNotes framework integrates Large Language Models (LLMs), Retrieval-Augmented Generation (RAG), and Automatic Speech Recognition (ASR) technologies to streamline medical note generation. The framework is designed to handle two main scenarios: generating SOAP (Subjective, Objective, Assessment, Plan) notes from medical conversations in real-time and retrieving relevant medical information based on user queries. The overall architecture of MediNotes is depicted in Figure 1, which demonstrates the flow from voice or text input to the generation and retrieval of structured medical documentation.

In Scenario 1, MediNotes operates in real-time, capturing both text and voice inputs during medical consultations. Using its integrated ASR system, the framework passively listens to conversations between physicians and patients, automatically transcribing spoken dialogue into structured SOAP (Subjective, Objective, Assessment, Plan) notes. The transcription is processed through an audio encoder, followed by fine-tuned LLM models, which refine and organize the content into detailed medical notes. This process significantly reduces the manual effort required for documentation, allowing physicians to focus more on patient care. The generated SOAP notes are stored in a vector database, enabling future retrieval for reference or further medical use.

In Scenario 2, MediNotes serves as a query-based retrieval system where both healthcare providers and patients can access medical information. Users can input their queries either through text or voice. The query is processed by an text encoder and sent to a Retrieval-Augmented Generation (RAG) model. The system searches the vector database for relevant content based on the user's query, retrieves contextual information, and formulates an appropriate response through the LLM. This scenario provides users with quick, accurate, and context-aware medical information, streamlining clinical workflows and improving the accessibility of essential data.

### *A. Audio Encoder*

The MediNotes framework incorporates an **audio encoder** coupled with a robust **Automatic Speech Recognition (ASR)** system to facilitate both real-time and non-real-time transcription of medical conversations. During interactions, the ASR system can capture voice inputs either in real-time as conversations occur or from pre-recorded audio files. This flexibility allows the system to handle both live clinical consultations and retrospective transcription of recorded sessions, providing physicians with a versatile solution for medical note generation.

At the core of the ASR system is the **Whisper-base** model, which ensures high-fidelity speech-to-text conversion. Additionally, the integration of **Pyannote-segmentation-3.0** for speaker diarization enables the framework to accurately identify and differentiate between the voices of multiple participants (e.g., physician and patient). The audio encoder captures and processes the input, whether in real-time through the user interface or from recorded audio, before the transcription is tokenized and passed through the LLM for further processing.

### *B. Dataset and Data Preprocessing*

To train and fine-tune the MediNotes model, we utilized the ACI-BENCH dataset [1], which contains 207 doctor-patient role-play dialogues, each averaging 1,302 tokens, with corresponding SOAP notes averaging 490 tokens. The dataset was split into three subsets: 67 dialogues for training, 20 for validation, and 120 for testing, further divided into three test sets. This dataset was selected for being the largest publicly available corpus of medical notes, encompassing outpatient scenarios, thereby improving the model's capacity to generalize across routine medical consultations.

The preprocessing phase involved several key steps:

- **Data Cleaning**: filling in missing values, removing outliers, and smoothing noisy data.
- **Text Normalization**: Removal of irrelevant characters and standardization of medical terms.
- **Tokenization**: Both dialogues and SOAP notes were tokenized using a SentencePiece tokenizer pre-trained on medical texts.

### *C. Generator Model*

Given the need for an advanced model capable of maintaining context in lengthy and complex medical dialogues, we selected **LLaMA3-8B** for its optimal balance between performance and resource efficiency. This model is specifically designed to handle long-range dependencies, making it well-suited for clinical dialogue summarization tasks. Additionally, **GEMMA-7B** and **Mistral-7B** were included for ablation studies to compare different model architectures and their effectiveness. These models excel in accuracy and adaptability, while also being highly efficient for fine-tuning using advanced techniques such as **QLoRA** and **PEFT**, ensuring they scale well in real-world healthcare applications.

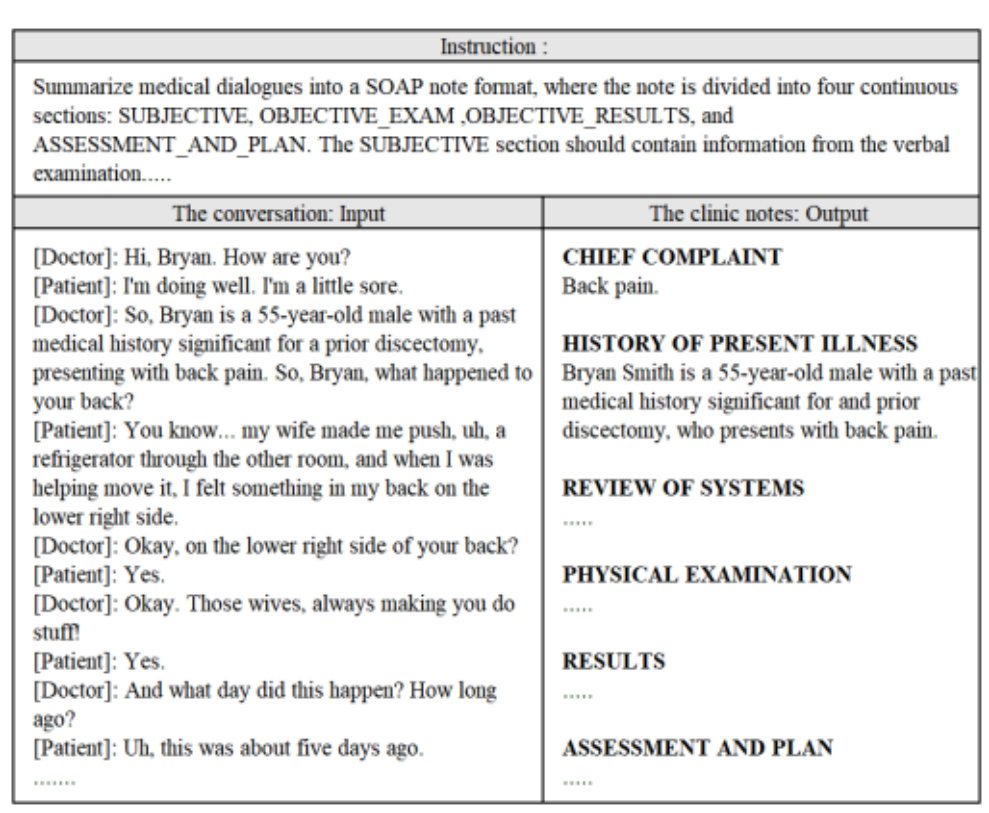

| Instruction : | |
|---|---|
| Summarize medical dialogues into a SOAP note format, where the note is divided into four continuous sections: SUBJECTIVE, OBJECTIVE_EXAM ,OBJECTIVE_RESULTS, and ASSESSMENT_AND_PLAN. The SUBJECTIVE section should contain information from the verbal examination..... | |
| The conversation: Input | The clinic notes: Output |
| [Doctor]: Hi, Bryan. How are you?<br>[Patient]: I'm doing well. I'm a little sore.<br>[Doctor]: So, Bryan is a 55-year-old male with a past medical history significant for a prior discectomy, presenting with back pain. So, Bryan, what happened to your back?<br>[Patient]: You know... my wife made me push, uh, a refrigerator through the other room, and when I was helping move it, I felt something in my back on the lower right side.<br>[Doctor]: Okay, on the lower right side of your back?<br>[Patient]: Yes.<br>[Doctor]: Okay. Those wives, always making you do stuff!<br>[Patient]: Yes.<br>[Doctor]: And what day did this happen? How long ago?<br>[Patient]: Uh, this was about five days ago.<br>....... | **CHIEF COMPLAINT**<br>Back pain.<br><br>**HISTORY OF PRESENT ILLNESS**<br>Bryan Smith is a 55-year-old male with a past medical history significant for and prior discectomy, who presents with back pain.<br><br>**REVIEW OF SYSTEMS**<br>.....<br><br>**PHYSICAL EXAMINATION**<br>.....<br><br>**RESULTS**<br>.....<br><br>**ASSESSMENT AND PLAN**<br>..... |

Fig 2: The Format of Instruction Tuning for Medical Dialogue to SOAP Note Conversion.

dff

To ensure that the LLM remains both accurate and computationally efficient, we employed two advanced fine-tuning techniques tailored for resource-constrained environments:

*a) Parameter-Efficient Fine-Tuning (PEFT)* [12]: PEFT fine-tunes only a small subset of critical model parameters, significantly reducing computational requirements. We set $r = 16$ and targeted key modules *(q_proj, k_proj, v_proj, o_proj, gate_proj, up_proj, down_proj)* responsible for attention and feed-forward operations. The *lora_alpha* was set to 16 to optimize scaling of the low-rank matrices, ensuring efficient training while maintaining high accuracy.

*b) Quantized Low-Rank Adaptation (QLoRA)* [13]: QLoRA reduces memory consumption by quantizing model parameters to 4 bits, allowing for efficient fine-tuning without compromising performance. This approach is ideal for healthcare settings where computational resources are limited, enabling the model to perform high-precision tasks, such as medical documentation, on standard hardware.

*c) Instruction Tuning*: This method ensures that the model generates coherent and structured SOAP notes by training it with specific task instructions. It improves the model's ability to organize conversations into distinct sections (Subjective, Objective, Assessment, and Plan).

### *D. Retriever*

The MediNotes framework utilizes Retrieval-Augmented Generation (RAG) to efficiently manage both medical note generation and information retrieval. This approach enhances the system's ability to handle text inputs from various sources, including user queries and converted voice conversations. The process begins by splitting the input text into smaller, manageable chunks using the RecursiveCharacterTextSplitter, which facilitates more efficient processing and retrieval.

Once the input text is segmented, the system transforms these text chunks into numerical representations, or embeddings, using open-source embedding models provided by Langchain. These embeddings capture the semantic meaning of the text, allowing the system to understand and retrieve relevant information based on context. The embeddings are stored in a vector database via the PGVector extension in an open-source PostgreSQL database. This vector store serves as the memory of the language model, allowing efficient retrieval of previously documented notes or other relevant content through vector similarity searches.

When a user submits a query—either via text input or through voice converted to text—the query is also converted into an embedding. This embedding is then used to search the vector store for related chunks of information. The relevant chunks retrieved from the vector database are combined with a predefined system prompt, and this augmented prompt is processed by the LLM to generate an accurate and contextually relevant response.

Additionally, SOAP notes generated by the system are encoded into **PDF format** and sent to both the chatbot and the RAG pipeline for storage in the vector database. This ensures that the notes are readily accessible for future retrieval, enhancing the system's capability to provide quick, accurate information when queried.

## IV. RESULTS AND DISCUSSION

### *A. Model Evaluation Results*

We employed a combination of quantitative metrics—ROUGE, BERTScore, and BLEURT for evaluating the MediNotes model along with qualitative assessments to evaluate the performance of the MediNotes framework. The quantitative metrics measured the accuracy, relevance, and coherence of the generated medical reports, while the qualitative assessments involved expert reviews by medical professionals. Our fine-tuned model, MediNotes LLM, based on LLaMA3-8B, was evaluated using the ACI-BENCH dataset across three test sets to ensure consistent and reliable results across diverse doctor-patient dialogues.

| | Rouge1 | Rouge2 | RougeL | RougeLsum | BERTScore -precision | BERTScore -recall | BERTScore -F1 | BLEURT |
|---|---|---|---|---|---|---|---|---|
| BART+FTSAMSum | 53.46 | 25.08 | 29.30 | 48.62 | 67.93 | **69.41** | 68.63 | 38.52 |
| GPT4o | 50.83 | 21.21 | **29.32** | 47.39 | 64.85 | 67.15 | 65.96 | 39.58 |
| Phi-3-mini-4k-instruct | 19.09 | 1.55 | 10.29 | 17.17 | 48.63 | 48.95 | 48.76 | 31.92 |
| Gemma-7b | 14.64 | 4.47 | 8.05 | 13.04 | 43.35 | 37.79 | 39.8 | 29.65 |
| Mistral-7b | 8.47 | 3.13 | 5.15 | 8.09 | 45.62 | 33.11 | 37.73 | 22.66 |
| Mistral-7b-instruct | 0.13 | 0 | 0.07 | 0.13 | 32.35 | 19.87 | 24.53 | 29.13 |
| Llama3-8B | 28.32 | 10.33 | 16.27 | 25.25 | 55.26 | 52.56 | 53.64 | 35.69 |
| Llama3-8b-instruct | 43.50 | 16.58 | 24.06 | 39.98 | 64.91 | 62.43 | 63.62 | 36.52 |
| Llama3-8B-FT (Our) | **56.16** | **28.91** | 29.30 | **51.6** | **70.56** | 69.31 | **70.34** | **41.05** |

Fig 3. Model Performance on Medical Note Generation Tasks (Test1 Evaluation)

| | Rouge1 | Rouge2 | RougeL | RougeLsum | BERTScore -precision | BERTScore -recall | BERTScore -F1 | BLEURT |
|---|---|---|---|---|---|---|---|---|
| BART+FTSAMSum | 52.08 | 24.37 | **28.84** | 47.16 | **67.61** | 68.74 | 68.16 | 37.29 |
| GPT4o | 50.95 | 21.37 | **28.84** | 47.93 | 65.05 | 66.84 | 65.9 | 39.46 |
| Phi-3-mini-4k-instruct | 18.40 | 1.25 | 10.1 | 16.77 | 48.85 | 48.82 | 48.81 | 31.31 |
| Gemma-7b | 15.7 | 4.85 | 8.62 | 14.38 | 44.91 | 38.04 | 40.59 | 30.78 |
| Mistral-7b | 1.13 | 0.28 | 0.67 | 1.11 | 40.85 | 26.31 | 31.79 | 21.07 |
| Mistral-7b-instruct | 46.18 | 19.3 | 26.6 | 41.66 | 66.74 | 62.41 | 64.49 | 39.48 |
| Llama3-8B | 33.4 | 13.67 | 20.4 | 30.7 | 58.04 | 55.12 | 56.42 | 36.73 |
| Llama3-8b-instruct | 40.54 | 15.37 | 22.3 | 37.76 | 64.58 | 61.23 | 62.8 | 35.49 |
| Llama3-8B-FT (Our) | **59.6** | **32.9** | 27.28 | **55.02** | 67.21 | **73.21** | **73.2** | **40.98** |

Fig 4. Model Performance on Medical Note Generation Tasks (Test2 Evaluation)

| | Rouge1 | Rouge2 | RougeL | RougeLsum | BERTScore -precision | BERTScore -recall | BERTScore -F1 | BLEURT |
|---|---|---|---|---|---|---|---|---|
| BART+FTSAMSum | 52.77 | 24.38 | 28.55 | 48.03 | 67.9 | **69.04** | 68.46 | 36.41 |
| GPT4o | 50.15 | 20.72 | 28.48 | 47.6 | 65.05 | 66.6 | 65.8 | 39.91 |
| Phi-3-mini-4k-instruct | 18.40 | 1.25 | 10.1 | 16.77 | 48.85 | 48.82 | 48.81 | 31.31 |
| Gemma-7b | 0 | 0 | 0 | 0 | 29.91 | 18.69 | 22.94 | 33.25 |
| Mistral-7b | 51.37 | 26.28 | 33.69 | 47.26 | 68.41 | 65.92 | 67.05 | 42.99 |
| Mistral-7b-instruct | 45.02 | 19.54 | 26.72 | 41.32 | 66.55 | 61.92 | 64.13 | 39.02 |
| Llama3-8B | 29.14 | 10.26 | 16.00 | 26.3 | 56 | 51.06 | 53.18 | 34.36 |
| Llama3-8b-instruct | 40.54 | 15.37 | 22.3 | 37.76 | 64.58 | 61.23 | 62.8 | 35.49 |
| Llama3-8B-FT (Our) | **58.91** | **31.74** | **37.34** | **54.91** | **72.98** | 72.59 | **72.75** | **41.43** |

Fig 5. Model Performance on Medical Note Generation Tasks (Test3 Evaluation)

The evaluation demonstrated that the MediNotes model consistently outperformed baseline models and other competitive architectures, including GPT4o, the best commercial model available at the time, and BART+FTSAMSsum, the top-performing model from the ACI-Bench dataset, across multiple key metrics. In terms of ROUGE (Recall-Oriented Understudy for Gisting Evaluation) measures the overlap between the generated and reference summaries, MediNotes showed the best performance in the 3 testing, achieving the highest Rouge1, Rouge2, and RougeLsum scores, demonstrating its ability to accurately capture relevant information from conversations. In Round 3, MediNotes scored 58.91 (Rouge1) and 54.91 (RougeLsum), significantly surpassing GPT4o and BART+FTSAMSsum.

The model also excelled in BERTScore, which evaluates the semantic similarity between the generated and reference summaries. Unlike ROUGE, which focuses on exact word overlap, BERTScore leverages contextual embeddings from a pre-trained BERT model to measure how well the meaning of the generated text matches the reference. MediNotes LLM achieved an F1 score of 73.2 in Round 2, reflecting its capability to generate semantically accurate and relevant summaries that align closely with the input dialogues. This was further supported by high BLEURT scores, which measure the quality of the generated text based on human-like quality and fluency. BLEURT evaluates how natural and coherent the output is, mimicking human judgment. MediNotes LLM consistently achieved scores in the 41.0-41.5 range for BLEURT, outperforming competitors like GPT4o and BART+FTSAMSsum.

While GPT4o and BART+FTSAMSsum performed well, they were generally outclassed by MediNotes LLM in most of the evaluated metrics. GPT4o showed strength in Rouge2 and BERTScore recall, but it lagged behind in overall coherence and precision. BART+FTSAMSsum, though performing well in BERTScore precision and recall, struggled with BLEURT and Rouge2 scores, indicating that its summaries were more formulaic and lacked the nuanced understanding needed for medical dialogue summarization.

Models like Phi-3-mini-4k-instruct, Gemma-7B, and Mistral-7B underperformed across most metrics, particularly in BLEURT and RougeL, suggesting that they were less effective in generating coherent and relevant summaries. These models struggled with capturing the complex medical language and dialogue structure required for the task.

In conclusion, the MediNotes model, based on LLaMA3-8B, stands out as the most effective model for medical dialogue summarization, offering superior performance in both quantitative and qualitative evaluations. Its ability to accurately generate medical reports from conversations with high precision and semantic relevance underscores its potential to improve clinical workflows and reduce the administrative burden on healthcare professionals.

### B. *Framework Evaluation Results*

A clinical evaluation was conducted in collaboration with the University of Chicago Medical Center to assess the performance of the MediNotes framework. Medical professionals reviewed the generated notes and data retrieval responses, evaluating them for accuracy, completeness, satisfaction, and usefulness. The study involved 10 doctors and 10 patients, each participating in 8 recorded conversations facilitated by the chatbot, as well as testing the chatbot with 8 query-based interactions.

- Accuracy: The percentage of correct information provided by the system compared to a gold standard.
- Completeness: The degree to which the system captures all necessary information.
- Satisfaction: User satisfaction levels based on feedback surveys.
- Usefulness: The perceived utility of the system in aiding users' tasks.

TABLE I: Quantitative Evaluation of the MediNotes Framework for Both Physicians and Patients Using Two Different Query Methods

| User Type | Query Type | Accuracy | Completeness | Satisfaction | Usefulness |
|---|---|---|---|---|---|
| Physicians | Conversation Recording | 75 | 60 | 70 | 85 |
| Physicians | Query-Based Test | 65 | 55 | 65 | 80 |
| Patients | Conversation Recording | 80 | 70 | 75 | 95 |
| Patients | Query-Based Test | 65 | 68 | 70 | 96 |
| Average | | 71 | 63 | 70 | 89 |

The findings revealed that 75% of the generated notes were deemed clinically usable without requiring manual corrections, while 60% of the notes achieved satisfactory levels of completeness. Furthermore, the satisfaction ratings of 70% reflected a positive reception from both physicians and patients. Notably, 89% of the evaluators indicated that the implementation of MediNotes in clinical settings could substantially reduce the administrative burden on healthcare providers, thereby enhancing operational efficiency and improving physician well-being.

### C. *Ablation Studies*

To better understand the impact of different model architectures and fine-tuning methods, an ablation study was performed to compare the performance of several models, including LLaMA3-8B, Mistral-7B, Gemma-7B, and Phi-3-mini-4k-instruct. This comparison aimed to isolate the effects of both model architecture and fine-tuning approaches, such as **QLoRA** and **PEFT**, on the overall performance of medical note generation tasks.

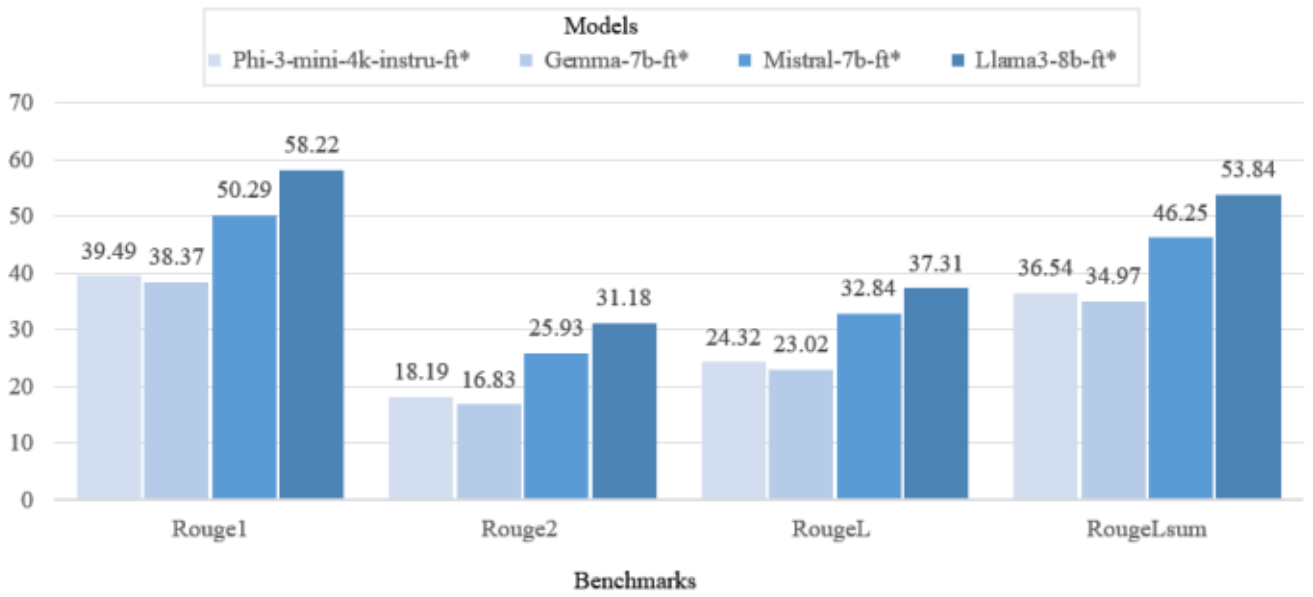

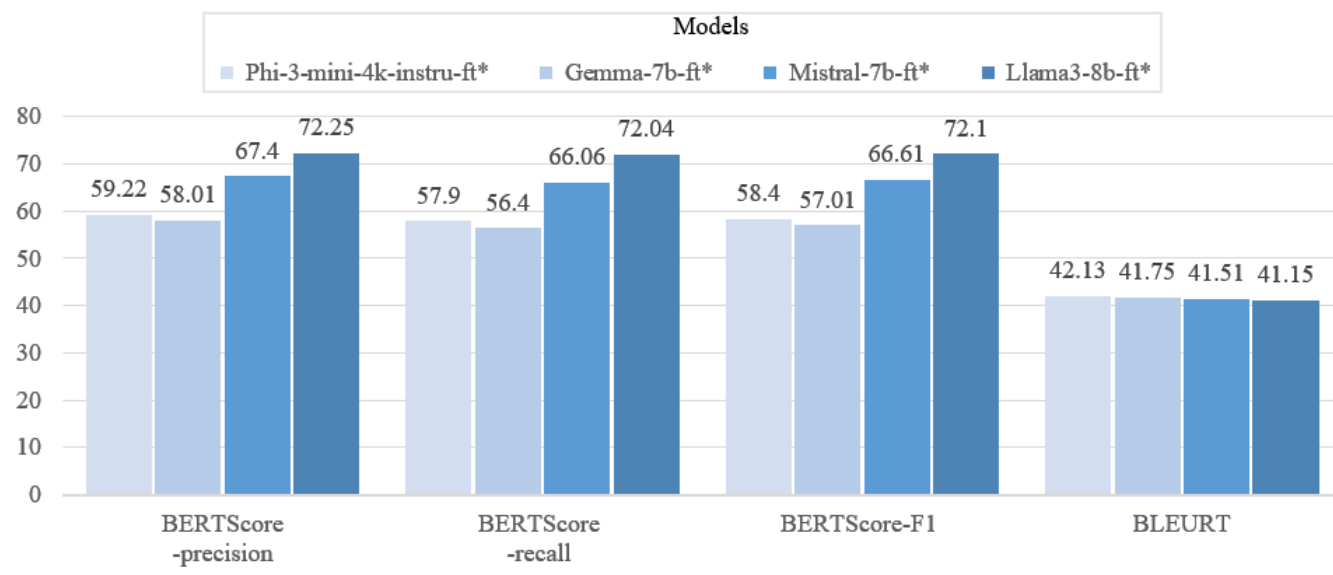

Fig 7. Performance Comparison of Models on ROUGE, BERTScore and BLEURT Metrics.

As depicted in Figure 7, each bar illustrates the performance of different models on several benchmarks**.** LLaMA3-8B-ft is the top-performing model across most metrics, demonstrating superior accuracy and coherence in generating medical summaries. It achieves the highest scores in ROUGE-1 (58.22), ROUGE-2 (31.18), and BERTScore-F1 (72.1), showing its ability to capture both individual words and contextual meaning effectively. With a strong ROUGE-Lsum score (53.84), it also excels in maintaining the overall structure of the summaries. Though slightly lower in BLEURT (41.15), it remains competitive in producing fluent and human-like text.

Mistral-7B-ft provides a solid balance between accuracy and fluency, making it a close competitor to LLaMA3-8B-ft. It scores well in ROUGE-1 (50.29), ROUGE-2 (25.93), and BERTScore-F1 (66.61), demonstrating its capability to generate relevant and structured content. Its BLEURT score of 41.51 is slightly higher than LLaMA3-8B-ft, indicating strong fluency and naturalness in the text. While it performs slightly lower overall, it remains a competitive option for accurate and readable summaries.

GEMMA-7B-ft excels in recall and fluency, though it lags in precision and content accuracy. It has a strong BLEURT score of 41.75, indicating that its generated text is highly fluent and human-like. However, its lower ROUGE-1 (38.37) and ROUGE-2 (16.83) scores show that it struggles to capture precise content. GEMMA-7B-ft's strength lies in producing broad, comprehensive summaries with a natural flow, making it suitable for tasks emphasizing readability. Phi-3-mini-4k-instru-ft stands out in producing the most fluent and natural summaries, reflected by its leading BLEURT score of 42.13. Despite its lower ROUGE-1 (39.49) and ROUGE-2 (18.19) scores, indicating weaker content accuracy, it still manages a decent BERTScore-Recall (57.9). This suggests it captures a broad range of content, though with less precision. Phi-3-mini-4k-instru-ft is best suited for tasks where fluency and readability are prioritized over detailed content accuracy.

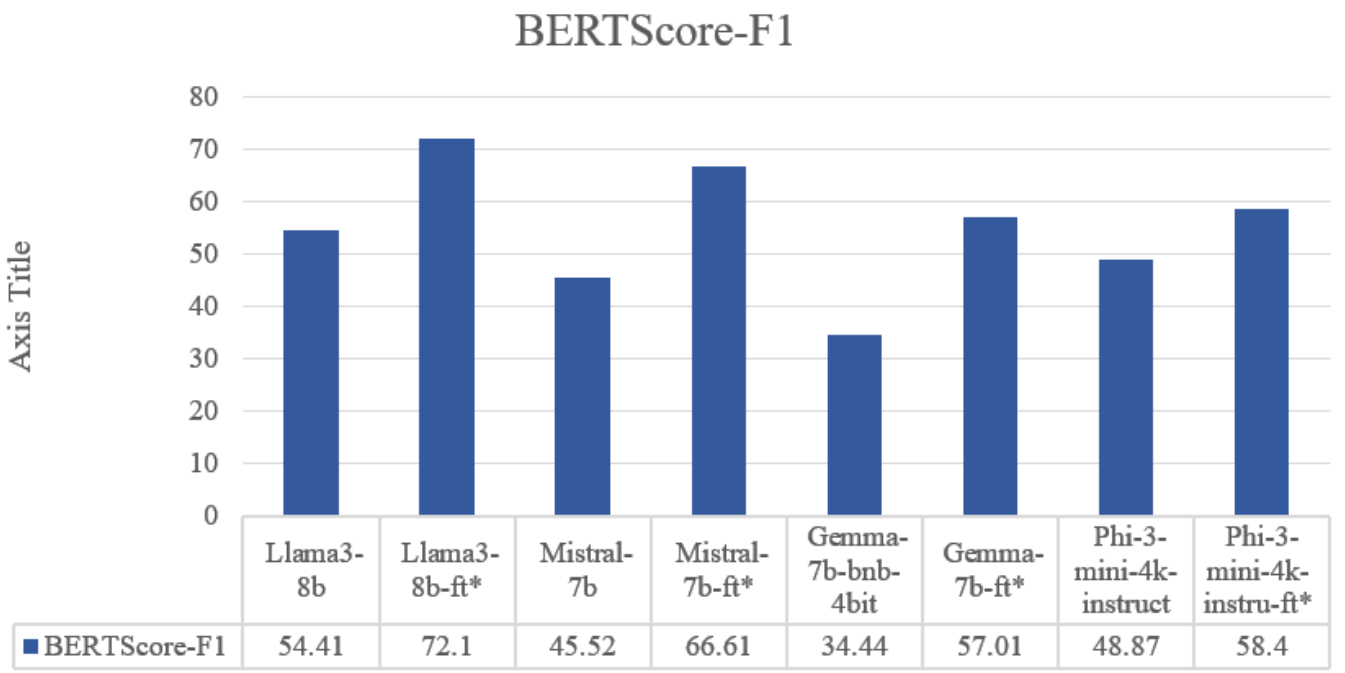

Fig 8. Impact of Instruction Tuning on ROUGE-1 Performance.

Figure 8 emphasizes the critical impact of fine-tuning, particularly instruction tuning, on the semantic relevance of generated summaries, as reflected by the BERTScore-F1 metric. The results clearly show that models with instruction tuning, such as LLaMA3-8B-ft (72.1) and Mistral-7B-ft (66.61), outperform their counterparts that either lack fine-tuning or use alternative configurations. Without instruction tuning, models like Mistral-7B (45.52) and Phi-3-mini-4k-instruct (48.87) see significant performance degradation, resulting in less accurate and structured outputs. These findings highlight that instruction tuning is pivotal in enhancing the quality and coherence of the generated text, as evidenced by the superior BERTScore-F1 scores of the fine-tuned models.

## V. DISCUSSION

### *A. Limitations*

The dataset used to train the MediNotes model faces significant limitations due to the sensitive nature of medical information and strict privacy regulations, such as HIPAA in the U.S. Acquiring real clinical data, especially recordings of patient-doctor conversations, is difficult because sharing such data risks compromising patient confidentiality. As a result, the dataset is relatively small and may rely on synthetic or anonymized data. Although this allows for the development of a functional model for clinical dialogue and note generation, it may not fully capture the diversity and complexity of real-world medical encounters, potentially affecting the model's generalization performance.

Additionally, the diversity of clinical language is a critical factor. While the dataset includes some variety in dialogue styles, expressions, and medical terminology from different healthcare providers and patients, it may lack the comprehensive scope required to train the model for a broad range of interactions. If the dataset is not sufficiently diverse, the model's ability to accurately understand and generate notes for varying dialogue patterns and medical language may be constrained. Therefore, to improve the model's adaptability and performance across diverse clinical environments, it is vital to incorporate a more representative and diverse dataset in future development efforts.

### *B. Application*

To MediNotes could be integrated into hospital Electronic Health Record (EHR) systems to automate the real-time generation of medical notes from patient-doctor conversations. By capturing spoken consultations via Automatic Speech

Recognition (ASR), it can generate structured notes like SOAP (Subjective, Objective, Assessment, Plan) that are instantly stored in the EHR. This reduces the time physicians spend on documentation, allowing them to focus more on patient care. In addition to generating notes, MediNotes can assist physicians by providing real-time access to critical patient data, such as previous lab results or diagnoses, supporting clinical decision-making. In telehealth settings, MediNotes can ensure efficient documentation and streamline the sharing of notes with other healthcare providers or patients.

Patients could also benefit from MediNotes by having access to their own medical records through the EHR system, enabling them to review their health information or treatment plans. However, the implementation of MediNotes comes with challenges. Ensuring data privacy and security in compliance with regulations like HIPAA is crucial, while addressing interoperability issues with different EHR systems will require tailored APIs and middleware. Additionally, adoption by healthcare staff may be slow without adequate training, making it necessary to demonstrate the efficiency and accuracy gains of the system to encourage widespread use.

### *C. Ethical considerations*

Ethical considerations surrounding the integration of AI in healthcare, particularly regarding privacy and model bias, are critical. The sensitive nature of medical data necessitates stringent privacy protections, ensuring that patient information remains confidential and secure. Moreover, AI system like MediNotes must be rigorously tested to prevent biases that could lead to unequal treatment or inaccuracies in medical documentation. Transparency in how these models function and make decisions is essential to building trust among both healthcare providers and patients. Accountability is also key, with physicians maintaining oversight to validate AI-generated notes, ensuring that AI supports rather than replaces clinical judgment, thus safeguarding the quality of care.

## VI. CONCLUSION

In conclusion, MediNotes presents a promising solution to alleviate the administrative burden on healthcare professionals by automating medical note generation through advanced AI techniques. By integrating large language models (LLMs), Retrieval-Augmented Generation (RAG), and Automatic Speech Recognition (ASR), it enables real-time documentation that improves the efficiency of clinical workflows. The framework has demonstrated superior performance in producing accurate, structured, and contextually relevant medical notes. However, while MediNotes has shown considerable potential, there is room for future improvements. Expanding the diversity of training datasets to better capture the range of clinical interactions and medical terminology will enhance the model's adaptability. Further advancements in data privacy, security, and model bias mitigation are essential to ensure ethical and safe integration into healthcare systems. Additionally, incorporating more sophisticated retrieval techniques and expanding interoperability with various EHR platforms could further boost its functionality, making MediNotes an even more effective tool in modern medical environments.

## REFERENCES

[1] T. D. Shanafelt, L. N. Dyrbye, C. Sinsky, et al., "Relationship Between Clerical Burden and Characteristics of the Electronic Environment With Physician Burnout and Professional Satisfaction," *Mayo Clin. Proc.*, vol. 91, no. 7, pp. 836-848, July 2016.

[2] J. Overhage and D. McCallie, "Too many clicks, too little time: Care provider perceptions of electronic health records," Int. J. Med. Inform., vol. 83, no. 3, pp. 206–219, Mar. 2014.

[3] A. Gupta, "Large language models in healthcare: Opportunities, challenges, and the future," Healthc. Sci., vol. 1, no. 2, pp. 100–106, 2023..

[4] J. Yim, H. Kim, J. Lee, and Y. Choi, "Clinical note generation from patient-doctor dialogues: The ACI-Bench corpus," Journal of Biomedical Informatics, vol. 134, p. 104144, 2023.

[5] H. Y. Leong, Y. Gao, S. Ji, Y. Zhang, and U. Pamuksuz, "Efficient Fine-Tuning of Large Language Models for Automated Medical Documentation," in *2024 4th IEEE International Conference on Digital Society and Intelligent Systems (DSInS)*, Sydney, Australia, 2024. doi: 10.1109/DSInS64146.2024.10992195.

[6] J. M. Kessels, "Patients' memory for medical information," *J. R. Soc. Med.*, vol. 96, no. 5, pp. 219–222, May 2003.

[7] P. Lewis *et al*., "Retrieval-augmented generation for knowledge-intensive NLP tasks," in *Advances in Neural Information Processing Systems 33*, 2020, pp. 9459–9474.

[8] J. Steinkamp, J. J. Kantrowitz, and S. Airan-Javia, "Prevalence and sources of duplicate information in the electronic medical record," *JAMA Network Open*, vol. 5, no. 9, pp. e2233348–e2233348, Sep. 2022.

[9] D. Yu and L. Deng, *Automatic speech recognition*, vol. 1. Berlin: Springer, 2016.

[10] L. Murray, D. Gopinath, M. Agrawal, and S. Horng, "MedKnowts: Unified Documentation and Information Retrieval for Electronic Health Records," presented at the Conf. on Human Factors in Computing Systems, 2020.

[11] W. Yim and M. Yetisgen-Yildiz, "Towards Automating Medical Scribing: Clinic Visit Dialogue2Note Sentence Alignment and Snippet Summarization," in *Proc. 2023 Int. Conf. on Biomedical Informatics*, 2023.

[12] N. Ding, Y. Qin, G. Yang, F. Wei, Z. Yang, Y. Su, S. Hu, Y. Chen, C. M. Chan, W. Chen, and J. Yi, "Parameter-efficient fine-tuning of large-scale pre-trained language models," Nature Machine Intelligence, vol. 5, no. 3, pp. 220–235, 2023.

[13] T. Dettmers, A. Pagnoni, A. Holtzman, and L. Zettlemoyer, "Qlora: Efficient finetuning of quantized LLMs," arXiv preprint arXiv:2403.08295, 2024.

[14] Y. Qiao, K. Li, J. Lin, R. Wei, C. Jiang, Y. Luo, and H. Yang, "Robust domain generalization for multi-modal object recognition," *arXiv preprint arXiv:2408.05831*, 2024.

[15] H. Y. Leong and Y. Wu, "Why Should Next-Gen LLM Multi-Agent Systems Move Beyond Fixed Architectures to Dynamic, Input-Driven Graphs?," SSRN Electronic Journal, 2024. doi: 10.2139/ssrn.5276004. Available: https://ssrn.com/abstract=5276004

[16] H. Zhang, B. Huang, Z. Li, X. Xiao, H. Y. Leong, Z. Zhang, X. Long, T. Wang, and H. Xu, "Sensitivity-LoRA: Low-Load Sensitivity-Based Fine-Tuning for Large Language Models," in Findings of the 2025 Conference on Empirical Methods in Natural Language Processing (EMNLP), 2025. doi: 10.48550/arXiv.2509.09119. Available: https://arxiv.org/abs/2509.09119

[17] J. Zhang, J. Gao, W. Ouyang, W. Zhu, and H. Y. Leong, "Time-LlaMA: Adapting Large Language Models for Time Series Modeling via Dynamic Low-rank Adaptation," in Proc. 63rd Annual Meeting of the Association for Computational Linguistics (ACL) – Student Research Workshop, 2025. doi: 10.18653/v1/2025.acl-srw.90. Available: https://aclanthology.org/2025.acl-srw.90/

[18] B. Wang, Y. Li, Q. Zhou, H. Y. Leong, T. Zhao, L. Ye, H. Deng, D. Luo, and N. Vasconcelos, "Do Vision Language Models Infer Human Intention Without Visual Perspective-Taking? Towards a Scalable 'One-Image-Probe-All' Dataset," in Proc. ICML Workshop on Assessing World Models, 2025. Available: https://openreview.net/forum?id=iekoq1rv80

[19] T. Qiu, J. Gao, J. Li, H. Y. Leong, and L. Zhang, "IntentVCNet: Bridging Spatio-Temporal Gaps for Intention-Oriented Controllable Video Captioning," in Proc. ACM Multimedia (MM), 2025. doi: 10.48550/arXiv.2507.18531. Available: https://arxiv.org/abs/2507.18531

[20] H. Y. Leong, Y. Gao, S. Ji, Y. Zhang, and U. Pamuksuz, "Efficient Fine-Tuning of Large Language Models for Automated Medical Documentation," in 2024 4th IEEE Int. Conf. on Digital Society and Intelligent Systems (DSInS), Sydney, Australia, 2024. doi: 10.1109/DSInS64146.2024.10992195. Available: https://ieeexplore.ieee.org/document/10992195

[21] C. Sun, H. Y. Leong, and L. Li, "Coarse-to-Fine Personalized LLM Impressions for Streamlined Radiology Reports," SSRN Electronic Journal, 2024. (ICML 2025 NewInML Workshop note). doi: 10.48550/arXiv.2508.15845. Available: https://arxiv.org/abs/2508.15845

[21] Y. Wang, J. Zhong, and R. Kumar, "A systematic review of machine learning applications in infectious disease prediction, diagnosis, and outbreak forecasting," Preprints, 2025.

[22] Y. Wang, Z. Wang, J. Zhong, D. Zhu, and W. Li, "Applications of Small Language Models in Medical Imaging Classification with a Focus on Prompt Strategies," arXiv preprint arXiv:2508.13378, 2025. doi: 10.48550/arXiv.2508.13378.

[23] Y. Wang, F.-L. Zhang, and N. A. Dodgson, "Target Scanpath-Guided 360-Degree Image Enhancement," in Proc. 39th AAAI Conf. on Artificial Intelligence (AAAI), AAAI Press, 2025. doi: 10.1609/aaai.v39i8.32881.

[25] Y. Wang, Z. Wang, J. Zhong, D. Zhu, and W. Li, "Applications of Small Language Models in Medical Imaging Classification with a Focus on Prompt Strategies," arXiv preprint arXiv:2508.13378, 2025. doi: 10.48550/arXiv.2508.13378.

[24] Y. Zheng, B. Zhong, Q. Liang, S. Zhang, G. Li, X. Li, and R. Ji, "Towards universal modal tracking with online dense temporal token learning," IEEE Trans. Pattern Anal. Mach. Intell., 2025.

[25] Y. Zheng, B. Zhong, Q. Liang, Z. Mo, S. Zhang, and X. Li, "ODTrack: Online dense temporal token learning for visual tracking," in Proc. AAAI Conf. Artif. Intell., 2024.

[26] Y. Zheng, B. Zhong, Q. Liang, N. Li, and S. Song, "Decoupled Spatio-Temporal Consistency Learning for Self-Supervised Tracking," in Proc. AAAI Conf. Artif. Intell., 2025.

[27] Y. Zheng, B. Zhong, Q. Liang, G. Li, R. Ji, and X. Li, "Toward unified token learning for vision-language tracking," IEEE Trans. Circuits Syst. Video Technol., 2023.

[30] Y. Zheng, B. Zhong, Q. Liang, Z. Tang, R. Ji, and X. Li, "Leveraging local and global cues for visual tracking via parallel interaction network," IEEE Trans. Circuits Syst. Video Technol., 2022.

[28] Q. Wang, J. Ke, H. Ye, Y. Lin, Y. Fu, J. Zhang, K. Keutzer, C. Xu, and Y. Chen, "Angles Don't Lie: Unlocking Training-Efficient RL Through the Model's Own Signals," arXiv preprint arXiv:2506.02281, 2025. doi: 10.48550/arXiv.2506.02281.

[29] Q. Wang, J. Ke, M. Tomizuka, K. Keutzer, and C. Xu, "Dobi-SVD: Differentiable SVD for LLM Compression and Some New Perspectives," in Proc. 13th Int. Conf. Learn. Representations (ICLR), 2025.

[30] Q. Wang, H. Ye, M.-Y. Chung, Y. Liu, Y. Lin, M. Kuo, M. Ma, J. Zhang, and Y. Chen, "CoreMatching: A Co-adaptive Sparse Inference Framework with Token and Neuron Pruning for Comprehensive Acceleration of Vision-Language Models," arXiv preprint arXiv:2505.19235, 2025. doi: 10.48550/arXiv.2505.19235.

[31] Q. Wang, S. Vahidian, H. Ye, J. Gu, J. Zhang, and Y. Chen, "CoreInfer: Accelerating Large Language Model Inference with Semantics-Inspired Adaptive Sparse Activation," arXiv preprint arXiv:2410.18311, 2024. doi: 10.48550/arXiv.2410.18311.

[32] Q. Wang, J. Ke, Z. Liang, and S. Zhang, "MathNAS: If Blocks Have a Role in Mathematical Architecture Design," Adv. Neural Inf. Process. Syst. (NeurIPS), vol. 36, pp. 47475–47486, 2023.

[33] Y. Wang, F.-L. Zhang, and N. A. Dodgson, "Scantd: 360° Scanpath Prediction Based on Time-Series Diffusion," in Proc. 32nd ACM Int. Conf. Multimedia (ACM MM), 2024. doi: 10.1145/3664647.3681315.

[34] Y. Wang, F.-L. Zhang, and N. A. Dodgson, "Target Scanpath-Guided 360-Degree Image Enhancement," in Proc. 39th AAAI Conf. Artif. Intell. (AAAI), 2025. doi: 10.1609/aaai.v39i8.32881.

[35] S. Zhou, Z. Tian, X. Chu, X. Zhang, B. Zhang, X. Lu, C. Feng, Z. Jie, P. Y. Chiang, and L. Ma, "FastPillars: A Deployment-Friendly Pillar-Based 3D Detector," arXiv preprint arXiv:2302.02367, 2023. doi: 10.48550/arXiv.2302.02367.

[36] S. Zhou, Z. Yuan, D. Yang, X. Hu, J. Qian, and Z. Zhao, "PillarHist: A Quantization-Aware Pillar Feature Encoder Based on Height-Aware Histogram," in Proc. IEEE/CVF Conf. Comput. Vis. Pattern Recognit. (CVPR), 2025, pp. 27336–27345. doi: 10.1109/CVPR2025.XXXX. *(DOI placeholder, replace when official DOI is available)*

[37] S. Zhou, Z. Yuan, D. Yang, Z. Zhao, X. Hu, Y. Shi, X. Lu, and Q. Wu, "Information Entropy Guided Height-Aware Histogram for Quantization-Friendly Pillar Feature Encoder," arXiv preprint arXiv:2405.18734, 2024. doi: 10.48550/arXiv.2405.18734.

[38] S. Zhou, L. Li, X. Zhang, B. Zhang, S. Bai, M. Sun, Z. Zhao, X. Lu, and X. Chu, "LiDAR-PTQ: Post-Training Quantization for Point Cloud 3D Object Detection," in Proc. 12th Int. Conf. Learn. Representations (ICLR), 2024. Available: https://openreview.net/forum?id=XXXX. *(replace with official link/DOI when available)*

[39] S. Zhou, J. Nie, Z. Zhao, Y. Cao, and X. Lu, "FocusTrack: One-Stage Focus-and-Suppress Framework for 3D Point Cloud Object Tracking," in Proc. 33rd ACM Int. Conf. Multimedia (ACM MM), 2025. doi: 10.1145/XXXX. *(replace with official DOI when available)*

[40] S. Zhou, S. Wang, Z. Yuan, M. Shi, Y. Shang, and D. Yang, "GSQ-Tuning: Group-Shared Exponents Integer in Fully Quantized Training for LLMs On-Device Fine-Tuning," in Findings Assoc. Comput. Linguistics: ACL 2025, Vienna, Austria, Jul. 2025, pp. 22971–22988. Available: https://aclanthology.org/2025.findings-acl.1178/

[41] X. Liang, Y. He, M. Tao, Y. Xia, J. Wang, T. Shi, J. Wang, and J. Yang, "CMAT: A Multi-Agent Collaboration Tuning Framework for Enhancing Small Language Models," arXiv preprint arXiv:2404.01663, 2024. doi: 10.48550/arXiv.2404.01663.

[42] J. Wang, Z. Zhang, Y. He, Z. Zhang, Y. Song, T. Shi, Y. Li, H. Xu, K. Wu, X. Yi, et al., "Enhancing Code LLMs with Reinforcement Learning in Code Generation: A Survey," arXiv preprint arXiv:2412.20367, 2024. doi: 10.48550/arXiv.2412.20367.

[43] Q. Yi, Y. He, J. Wang, X. Song, S. Qian, X. Yuan, L. Sun, Y. Xin, J. Tang, K. Li, et al., "SCORE: Story Coherence and Retrieval Enhancement for AI Narratives," arXiv preprint arXiv:2503.23512, 2025. doi: 10.48550/arXiv.2503.23512.

[44] W. Liu, J. Xu, F. Yu, Y. Lin, K. Ji, W. Chen, Y. Xu, Y. Wang, L. Shang, and B. Wang, "QFFT: Question-Free Fine-Tuning for Adaptive Reasoning," arXiv preprint arXiv:2506.12860, 2025. doi: 10.48550/arXiv.2506.12860.

[45] W. Liu, J. Chen, K. Ji, L. Zhou, W. Chen, and B. Wang, "RAG-Instruct: Boosting LLMs with Diverse Retrieval-Augmented Instructions," arXiv preprint arXiv:2501.00353, 2024. doi: 10.48550/arXiv.2501.00353.

[46] Y. Zhang, X. Liu, R. Tao, Q. Chen, H. Fei, W. Che, and L. Qin, "ViTCoT: Video-Text Interleaved Chain-of-Thought for Boosting Video Understanding in Large Language Models," arXiv preprint arXiv:2507.09876, 2025. doi: 10.48550/arXiv.2507.09876.

[47] Y. Zhang, X. Liu, R. Zhou, Q. Chen, H. Fei, W. Lu, and L. Qin, "CCHall: A Novel Benchmark for Joint Cross-Lingual and Cross-Modal Hallucinations Detection in Large Language Models," arXiv preprint arXiv:2505.19108, 2025. doi: 10.48550/arXiv.2505.19108.

[48] S. Zeng, X. Chang, M. Xie, X. Liu, Y. Bai, Z. Pan, M. Xu, and X. Wei, "FutureSightDrive: Thinking Visually with Spatio-Temporal CoT for Autonomous Driving," arXiv preprint arXiv:2505.17685, 2025. doi: 10.48550/arXiv.2505.17685.

[49] C. Xue, B. Zhong, Q. Liang, Y. Zheng, N. Li, Y. Xue, and S. Song, "Similarity-Guided Layer-Adaptive Vision Transformer for UAV Tracking," in Proc. IEEE/CVF Conf. Comput. Vis. Pattern Recognit. (CVPR), 2025, pp. 6730–6740. Doi: 10.48550/arXiv.2503.06625

[50] C. Xue, B. Zhong, Q. Liang, H. Xia, and S. Song, "Unifying Motion and Appearance Cues for Visual Tracking via Shared Queries," IEEE Trans. Circuits Syst. Video Technol., 2024. doi: 10.1109/TCSVT.2024.3486347.

[51] B. Lin, J. Zheng, C. Xue, L. Fu, Y. Li, and Q. Shen, "Motion-Aware Correlation Filter-Based Object Tracking in Satellite Videos," IEEE Trans. Geosci. Remote Sens., vol. 62, pp. 1–13, 2024. doi: 10.1109/TGRS.2024.3350988

[52] Y. Wu, X. Liu, C. Zhao, and X. Wu, "Prompt-Guided Dual Latent Steering for Inversion Problems," arXiv preprint arXiv:2509.18619, 2025. doi: 10.48550/arXiv.2509.18619. (Accepted at DICTA 2025, oral)

[53] X. Liu, Y. Lu, X. Wang, and X. Wu, "Training-Free Multi-Style Fusion Through Reference-Based Adaptive Modulation," arXiv preprint arXiv:2509.18602, 2025. doi: 10.48550/arXiv.2509.18602. (Accepted at ACPR 2025, oral)